\documentclass[amsmath,amssymb,twocolumn,nofootinbib, superscriptaddress]{revtex4-1}

\usepackage[utf8]{inputenc}
\usepackage{times,bbm}
\usepackage{amsfonts}
\usepackage{xcolor}
\usepackage{mathtools}

\usepackage{graphicx}
\usepackage{float}
\usepackage{csquotes}

\definecolor{newcolor}{RGB}{0,0,0}
\newcommand{\new}[1]{{\color{newcolor}#1}}
\usepackage{hyperref}

\hypersetup{
	breaklinks,
	colorlinks,
	linkcolor=gray,
	citecolor=gray,
	urlcolor=gray,
}

\usepackage{dsfont}

\definecolor{jens}{rgb}{0,0,0}
\definecolor{dante}{rgb}{0,0,0}
\definecolor{cadmiumgreen}{rgb}{0.0, 0.42, 0.24}
\newcommand{\je}[1]{{\color{jens}#1}}
\newcommand{\dmk}[1]{{\color{dante}#1}}

\definecolor{augustine}{rgb}{0,0.5,1}

\begin{document}

%\title{Quantum time crystals with programmable disorder in higher dimensions}
\title{Quantum time crystals with programmable disorder in higher dimensions}

\author{A. Kshetrimayum}

\affiliation{\mbox{Helmholtz Center Berlin, 14109 Berlin, Germany}}
\affiliation{Dahlem Center for Complex Quantum Systems, Freie Universit{\"a}t Berlin, 14195 Berlin, Germany}

\author{M. Goihl}

\affiliation{Dahlem Center for Complex Quantum Systems, Freie Universit{\"a}t Berlin, 14195 Berlin, Germany}

\author{D. M. Kennes}
\affiliation{Institut f\"ur Theorie der Statistischen Physik, RWTH Aachen, 
52056 Aachen, Germany and JARA - Fundamentals of Future Information Technology}
\affiliation{Max Planck Institute for the Structure and Dynamics of Matter and Center for Free Electron Laser Science, 22761 Hamburg, Germany}
\author{J. Eisert}

\affiliation{Dahlem Center for Complex Quantum Systems, Freie Universit{\"a}t Berlin, 14195 Berlin, Germany}
\affiliation{\mbox{Helmholtz Center Berlin, 14109 Berlin, Germany}}

\date{\today}

\begin{abstract}
We present fresh evidence for the presence of discrete quantum time crystals in two spatial dimensions. Discrete time crystals are intricate quantum systems that break discrete time translation symmetry in driven quantum many-body systems undergoing non-equilibrium dynamics. They are stabilized by many-body localization arising from disorder. We directly target the thermodynamic limit using instances of infinite tensor network states and implement disorder in a translationally invariant setting by introducing auxiliary systems at each site. We discuss how such disorder can be realized in programmable quantum simulators:
\new{This gives rise to the interesting situation in which a classical tensor network simulation can contribute to devising a blueprint of a quantum simulator featuring pre-thermal time crystalline
dynamics, one that will yet ultimately have to be built in order to  explore the stability of this phase of matter for long times.}
%probing the stability of this phase of matter that are quantum technological devices intermediate between analog quantum simulators and quantum computers, in a physical situation where classical tensor network methods are available only for short and intermediate times.
\end{abstract}
\maketitle

\section{Introduction}
A \emph{time crystal} is a periodic structure that not only repeats itself
periodically in space, as regular crystals do, but also in time 
\cite{SondhiReview,doi:10.1146/annurev-conmatphys-031119-050658}. The idea
of a crystal in space and
time has been proposed by 
Franck Wilczek \cite{PhysRevLett.109.160401} %who has envisioned to realizing a time crystal
using superconducting rings and persistent currents. Not much later, it turned out that serious obstructions need to be overcome: Ground states of 
local Hamiltonians can provably not host a spontaneous breaking of 
time-translation symmetry \cite{PhysRevLett.114.251603,2ndNoGo}. So
achieving a quantum time crystal becomes possible only in two special scenarios: (i) using a non-local Hamiltonian or (ii) resorting to non-equilibrium dynamics. %either the assumption of a local interaction
%or that of equilibrium needs to be forfeited.
Indeed, one can construct Hamiltonian models the ground states of which feature quantum time crystals, albeit at the price of an intricate interaction pattern that is presumably
excessively challenging to achieve even in highly engineered quantum systems \cite{PhysRevLett.123.210602}. 

Or, as an alternative route much more 
amenable to experimental realization, one can resort to settings of \emph{non-equilibrium physics} 
\cite{PhysRevLett.116.250401,PhysRevB.93.245145,PhysRevB.93.245146,1408.5148,ngupta_Silva_Vengalattore_2011}. 
Specifically, systems undergoing a 
time-periodic driving in one spatial dimension have been identified as suitable candidates to exhibit
discrete time translation symmetry breaking
\cite{PhysRevLett.116.250401,PhysRevA.91.033617,PhysRevLett.118.030401,PhysRevB.93.245146,PhysRevLett.117.090402,PhysRevLett.118.030401,0706.0212,SondhiReview,doi:10.1146/annurev-conmatphys-031119-050658},
referred to as \emph{discrete time crystals}, going back to seminal work of
Khemani, Lazarides, Moessner, and Sondhi
\cite{PhysRevLett.116.250401}.
Indeed, this idea has led to 
the experimental observation of time crystals, both in one-dimensional 
systems of trapped ions using a programmable potential
\cite{ZhangNature2017}
and in large settings of dipolar spin impurities in diamond at room-temperature
\cite{LukinTimeCrystal}. A further obstacle that arises along the way 
in  such periodically driven physical systems,
can be largely overcome. To avoid heating due to the driving, disorder comes
to rescue, giving rise to \emph{many-body localization}
\cite{huse2014phenomenology,RevModPhys.91.021001}
that % allows for equilibration and non-equilibrium
%dynamics, but 
prevents thermalization 
\cite{Zhang2017,MonroeTimeCrystal,LukinTimeCrystal,KennesFloquet},
 either for long, as in pre-thermal, or all times
\cite{Sondhi,KhemaniNISQTC}.
That is to say, by suitably combining disorder and periodic driving, the realization of this intricate state of matter is conceivable in synthetic
quantum devices. Indeed, experimental realizations of such discrete time crystals can be viewed as instances of
\emph{dynamical quantum simulations} 
\cite{CiracZollerSimulation,BlochSimulation,Trotzky}
\dmk{realizing} quantum technological devices \cite{RoadMap}.
\dmk{In this sense} they can be seen as
physical systems that allow to probe intriguing features of interacting quantum many-body
systems under precisely controlled conditions.

This state of affairs, \dmk{discussed in the previous paragraph}, is most interesting in situations in which state-of-the art classical simulation techniques can provide strong
evidence of the existence of discrete time crystals.
At the same time a full simulation in all aspects is
out of reach with present classical simulation tools
\new{(and in fact presumably 
for all 
classical simulation tools, 
as there are complexity-theoretic obstructions to this effect
\cite{Vollbrecht}).}
This points to the \new{exciting} 
direction of exploring discrete time
crystals in higher dimensions, settings that are conceivable
in \emph{programmable quantum simulators}
\cite{GoogleSupremacy,MonroeTimeCrystal,FlammiaProgrammable,Browaeys,Dimitris,ProbingQuantumSimulator,2dMBLScience,2DIons},
but for which the best known classical codes can just so keep track of the relevant 
features in time. \new{Indeed, this 
state of affairs} gives rise to the interesting situation that one can build trust in the functioning of a quantum simulator in relevant regimes. At the same time, there is scope for explorations outside the classically computationally accessible realm: After all, the full quantum simulation is barely beyond the boundary of what can be assessed with classical computers. And along these lines, e.g.,
discriminate pre-thermal behaviour for long times from genuine infinite-time time crystals \cite{SondhiReview,doi:10.1146/annurev-conmatphys-031119-050658}.

%It is the purpose of this work to present such a state-of-the art classical simulation that provides 
In this work, we provide evidence for
the existence of discrete time crystals in two spatial dimensions.
%supported for long times. 
This is possible by resorting to tensor network techniques such as \emph{projected entangled pair states (PEPS)} \cite{PEPSOld,VerstraeteBig,SierraOld,NishinoOld,Orus-AnnPhys-2014},
\new{here applied to and pushed further to a regime involving interactions, disorder and a time-periodic drive in two spatial dimensions}. Unlike other numerical  techniques, these tools are built to precisely capture genuine quantum correlations \cite{AreaReview} and can directly target the thermodynamic limit (iPEPS) \cite{iPEPSOld,SierraOld,NishinoOld}, thus overcoming finite size effects that are often encountered in classical simulations of quantum many-body systems. Targeting the thermodynamic limit is quite important especially in the context of diagnosing \new{time crystals} as they are only meaningful in the infinite-size limit
~\cite{SondhiReview}. For these reasons, iPEPS have been successfully used in the past to calculate ground states of challenging condensed matter problems \cite{Xiangkhaf,ThibautspinS,Corboz2DHubbard,KshetrimayumkagoXXZ,CorboztJ,CorbozMaterial,Kshetrimayummaterial}, thermal states \cite{piotr2012,piotr2015,piotr2016,Kshetrimayumthermal} as well as non-equilibrium steady states \cite{Kshetrimayum}. Tensor networks, in general, has the limitation that it cannot handle large entanglement build-up associated with studying time evolution and therefore, cannot go to very long times as this would require an exponentially large bond dimensions. 

\new{In fact, the challenge to reach longer times for tensor network algorithms is rooted in fundamental complexity theoretic obstructions, arising from
considerations in quantum information theory: The upshot is that
the development of efficient classical simulation methods for all time-evolving local Hamiltonians is implausible because such a time evolution turns out to be ultimately as powerful as a full quantum computer. In principle, any quantum algorithm could be run on a local Hamiltonian quantum system undergoing such a time evolution.
In more technical terms, 
local Hamiltonian time evolution is
 {\tt BQP} complete
~\cite{Vollbrecht} and therefore, a universal classical simulation of local Hamiltonian evolution is not expected to} be possible for all times. Despite this, the application of tensor networks has been extended to the challenging realm of keeping track of the time evolution in two-dimensional interacting systems \cite{piotrevolution,Claudiusevolution,Augustine2DMBL,Claudiuslocalquench,Kennes2018} to unprecedented times recently. 

Equipped with this machinery, we take on the problem of one of the most intricate quantum phases of matter: \new{These are} quantum time crystals in two dimensions. \new{To be concrete}, we do so by building upon recent efforts to capture time evolution in many-body localized systems in two spatial
dimensions \cite{Augustine2DMBL,Clausiusevolution}. Indeed, the stability of many-body localization in two spatial dimensions is an open and challenging topic due to the argument of an avalanche mechanism caused by the presence of rare ergodic bubbles~\cite{RoeckImbrie,deroeck2017} and with existing tensor network methods, we may never be able to predict the stability of such a phase at infinite times. 

In fact, \new{again} it is clear that any such classical simulation %must come to an end in time. 
\new{must face obstructions when aiming at assessing the behaviour of the system at long times. That said, by substantial computational effort and suitable algorithm development,}
on the practical side,  we can push things to show the existence of the phenomenon and to make a strong case for many-body localization in \new{two spatial 
dimensions}, \new{within the realm of finite times that can be assessed}. In fact, recently, there has been more evidence on many-body localization in \new{two spatial dimensions}~\cite{WahlNature2019,WahlPRB2020,Kennes2018,Alet2dMBLPRR2020,AletMBL2d2020} using a variety of numerical techniques. In Ref.~\cite{Augustine2DMBL}, we have shown the existence of \new{many-body localization} in 
\new{two spatial dimensions} for the longest times available to us based on the particle imbalance and the growth of local R{\'e}nyi entropies with programmable discrete disorder. We found that taking sufficient number of disorder levels as well as disorder strength is required to achieve localization, albeit for the available times. Besides, as we have also found in extensive numerical work, as the instability has its origin in thermal bubbles, the probability measure underlying the disorder matters significantly, a situation that can be easily accommodated with programmable disorder, say, in programmable quantum simulators as we have them in mind here. Our current work is built-up on this set up.
To achieve
our goal, we employ the evolution under a time dependent Floquet Hamiltonian for the first time within an iPEPS framework. 

Our advances are basically threefold: (i) We establish 
(at least pre-thermal)
\new{the presence of} 
time-crystalline behavior in higher than one spatial dimension
in unprecedented regimes over other known 
methods, 
thereby (ii) also demonstrating that tensor networks
are able to make meaningful predictions in this regime
\new{in which interactions, disorder and time-dependent drives come together}, and (iii)
we signify the power of programmable disorder to stabilize such behavior at longer times, \new{thereby providing further impetus for such quantum simulators from the perspective of quantum information science to assess the stability of intricate quantum phases of matter.}

\section{Model}

\new{In what follows,}
we consider a piece-wise constant
time dependent family of 
Floquet Hamiltonians $t\mapsto H_F(t)$ as 
\begin{equation}
\begin{split}
    H_F (t)= 
    \begin{cases}
     & H_{\rm MBL} \ \text{if} \ \ t  \in |0,T/2), \\
     & H_S \ \text{if} \ \ t  \in |T/2,T )
    \label{Hmain}.
\end{cases}
\end{split}
\end{equation}
Specifically, the time evolution operator advancing the system by one period is taken as $U=e^{-i H_ S T/2} e^{-i H_{\rm MBL} T/2}$, with $ H_{\rm MBL}$ being a static many-body localizing Hamiltonian described below.  
This entails that the total Hamiltonian is time periodic with period $T$, where within one period of $t\in [0,T)$, first only $H_{\rm MBL}$ is active for times $t\in [0,T/2)$, while for the second part of the period $t\in [T/2,T)$ only $H_S$ acts on the systems.
$H_S$ is a spin-flip operator that flips each spin in the system at suitable moments in time. 
We choose $H_S$ to be a spin flip operator with some periodicity $T$ as in Ref.~\cite{PhysRevLett.118.030401} as
\begin{equation}
    H_S = (2\pi/T-2\epsilon) \sum_i S^x_i.
\end{equation}
The action of this Hamiltonian is to
flip each spin in the system in the \new{ spin-z} basis $\{|\uparrow\rangle,| \downarrow \rangle\}$ at suitable times. \new{ $S^x$ is the spin-1/2 operator in the spin-x basis acting on each site, working in units $\hbar=1$.} The frequency of this flipping depends on the time period $T>0$. The real constant parameter $\epsilon$ determines the deviation from a perfect flip $( |\uparrow\rangle \mapsto |\downarrow \rangle )$ and vice versa: The persistence of a time crystal for non-zero values reflects a macroscopic rigidity since there is a periodic response in a measurable observable even when the driving is not synchronized. A discrete time crystal is reflected
by a response that is periodic in \emph{integer multiples} of the driving period (here, period doubling given our specific setting). 
For the $H_{\rm MBL}$ contribution to the Hamiltonian, we choose a Heisenberg model on a square lattice with 
\emph{discrete} disorder as
\begin{equation}
    H_{\rm MBL} = J \sum_{\langle i,j \rangle}\Vec{S}_i \cdot \Vec{S}_j + \sum_i h_i S_i^z,
    \label{Hmbl}
\end{equation}
where $h_i$ is drawn from a discrete uniform distribution with $d_a$ levels between $-h/2$ and $h/2$ {(the choice of naming the integer number of levels $d_a$ will become apparent below)}.
The above Hamiltonian 
is not only amenable to quantum simulation in
programmable quantum simulators: It can at the same
time be implemented in a perfectly
translationally invariant fashion using iPEPS \cite{Augustine2DMBL}{, by {appending} to   every local physical  spin-1/2 (denoted by sub-index p)  an auxiliary  one (denoted by sub-index a) with local Hilbert space dimension $d_a$. This {\it a posteriori} explains the choice of variable name above}. We have found
evidence of  localization based on the particle number very recently at sufficiently strong and enough levels of disorder. The disordered Heisenberg
model is a paradigmatic model for many-body localization, a feature that is preserved under
discrete disorder. In our tensor network simulation, briefly speaking, we start off from an initial state which is a tensor product of the physical initial state (local dimension $d_p=2$) and an auxiliary initial state (local dimension $d_a$), i.e., a state vector $|\Psi_0\rangle = |\psi_{p_0} \rangle \otimes |\psi_{a_0} \rangle $ where $|\psi_{a_0} \rangle = |\cdots ,+,+,+,+ \cdots \rangle$ and \new{ $|+\rangle = d_a^{-1/2}
(\sum_{s=1}^{d_a}|s\rangle)$, $s$ being the integer reflecting the allowed spin state determined by $d_a$}. 
  %\dmk{this sentence is still a little implicit: S is not defined s is defined only in a language that at least I am somewhat unused to.} 
The initial physical state vector $|\psi_{p_0} \rangle$ can be freely chosen. In this work, we make use of two different initial state vectors: $|\psi_{p_0} \rangle = |\cdots, \uparrow, \downarrow, \uparrow, \downarrow, \cdots\rangle$ (arranged in a checkerboard pattern, shown as a cartoon on the left side of the top panel of 
Fig.\ \ref{nodisorder} and others) and $|\psi_{p_0} \rangle = |\cdots, \uparrow,\uparrow, \uparrow, \uparrow, \cdots\rangle$, referred to as the N{\'e}el and spin polarized state vectors, respectively in the following. These initial states are readily experimentally accessible and can be viewed as two limits of the entire state space of random product states. The dynamics of a given initial random product states, e.g., with respect to its decay (of the expectation values of the observables), will behave somewhere ``in between'' those two extreme configurations. Another reason for choosing these initial states is that because we are directly targeting the thermodynamic limit, we need to assume some translational invariance with a fixed unit cell. In our setting, we work with a two-site unit cell arranged in a checkerboard patter (shown in Fig.~\ref{ipepsTC}). \new{The two initial states can be written exactly as an iPEPS with bond dimension one.}

The dynamics generated by Hamiltonian \eqref{Hmbl} including the \emph{disorder average} can be implemented by introducing $S^zS^z$ coupling in the Hamiltonian between the physical and the auxiliary sites. This term ensures that
the exact disorder average is recovered for all times and all disorder distributions. The modified form of \eqref{Hmbl} then takes the form
\begin{equation}
    H_{\rm MBL} = J\sum_{\langle i,j \rangle}\left( S^x_{i_p}S^x_{j_p} + S^y_{i_p}S^y_{j_p} + S^z_{i_p}S^z_{j_p} \right)
    + h\sum_i S^z_{i_p}S^z_{i_a}
    \label{Hanc} ,
 \end{equation}
 where the subscripts $p$ and $a$ is used to denote the physical and the auxiliary sites. The disorder average of all the possible configurations is already taken in one simulation while computing the expectation values of the observables (discussed in more detail in the method section of the paper). This is another advantage of our setting because we do not need to take multiple shots of disorder configuration and average them.

To monitor the time crystalline behavior we will extend the notion of  {\it long-ranged 
order}   
captured by order parameters from 
{\it equal-time correlations in space} 
to general correlations in space and
time 
\begin{equation}
C_{i,i'}^{OO}(t,t')=\lim_{|L|\rightarrow\infty}
\langle O(i,t) O(i',t')\rangle \neq 0 
\end{equation}
%\begin{equation}
%C_t^{OO}(i,i')=\lim_{|L|\rightarrow\infty}%\langle O(i,t) O(i',t)\rangle \neq 0 
%\end{equation}
where we put specific emphasis on 
\emph{equal-space correlations in time}, 
abbreviated as
\begin{equation}
C_i^{OO}(t,t')=\lim_{|L|\rightarrow\infty}\langle O(i,t) O(i,t')\rangle = f(t,t'),
\label{eqspacecorrelator}
\end{equation}
for times $t,t'\geq0$. For our purposes,
we call a system a time-crystal if,
robust to perturbations $\epsilon\neq 0$
(to avoid trivial situations as 
considered in Ref.\ \cite{PhysRevLett.114.251603}
being called a time crystal), $f(t,t')$ shows a non-trivial (ordered) long time $t\gg t'$ behaviour that breaks the discrete time-translation symmetry of the underlying drive (here by period doubling
or more generally taking integer values).\footnote{It 
is interesting to note
that a \emph{Magnus expansion} in terms of the
periodicity $T$ will lead to a perturbative
series of Hamiltonians each of which 
is captured by the no-go-theorem of Refs.\
\cite{PhysRevLett.114.251603,2ndNoGo}, so that in 
such a description, all orders must be considered.} 
We will concentrate on $O=S^z_i$ and $t'=0$ and write the corresponding correlation function
\new{as} $C_i^{S^zS^z}(t,0)=C_i^{zz}(t)$ \new{in} a shorthand notation. The expectation values of these correlators are computed using 
\new{an instance of a}
\emph{corner transfer matrix renormalization group} (CTMRG) \cite{ctmnishino1996,ctmnishino1997,ctmroman2009,ctmroman2012}
(\new{see also the methods section for the techniques being used)}. %This is a renormalization scheme devised for contracting the infinite two dimensional tensor network, this step being the main bottleneck in iPEPS simulations, setting it apart from the one-dimensional matrix product state calculations. We use an environment bond dimension of up to $D_{\rm CTM}=D^2$, where $D$ is the bond dimension of the iPEPS.}

\new{In our subsequent detailed discussion}, we specifically
consider the following different cases
that seem particularly important and insightful: These are (i) 
uncoupled spins without disorder for $J=0$, $h=0$.  Then, (ii)
coupled spins without disorder, $J=1$, $h=0$.  Finally, (iii) 
we investigate coupled spins with strong disorder, 
$J=1$, $h=100$ for different levels of disorder, i.e., 
$d_a=2$ and $d_a=5$ where $d_a$ corresponds to the size of the local Hilbert space of the auxiliary systems. 

\section{Results} 
In all the cases discussed above, we choose $T=0.1$ where $T$ is the periodicity of the Floquet cycle. We consider a perfect flip $\epsilon=0$ and an imperfect flip $\epsilon=0.5$ in order to check the robustness of the time crystal in all the cases.% We use a Trotter step of $\delta t=0.005$ in the time evolution and go up to bond dimension $D=5$ of the iPEPS due to the large physical dimension involved $d=d_p\cdot d_a=4$ and $d=d_p\cdot d_a=10$. 
To make sure that our results are not an artefact of the finite bond dimension, we make sure that the expectation values are consistent for the two best available bond dimensions we can afford, i.e.,
$D=4,5$ as our stopping criterion in time, identifying the time when the iPEPS ansatz can no longer
\new{faithfully}
accommodate the entanglement growth (discussed in more detail in the methods section).
We \new{largely focus on} these cases in order to understand the role of the various parts of the Hamiltonian and to establish which parameters of the Hamiltonian needs to be tuned in actual quantum simulators in order to stabilize or realize the time crystal.  

\begin{figure}
 \includegraphics[width=0.48\textwidth]{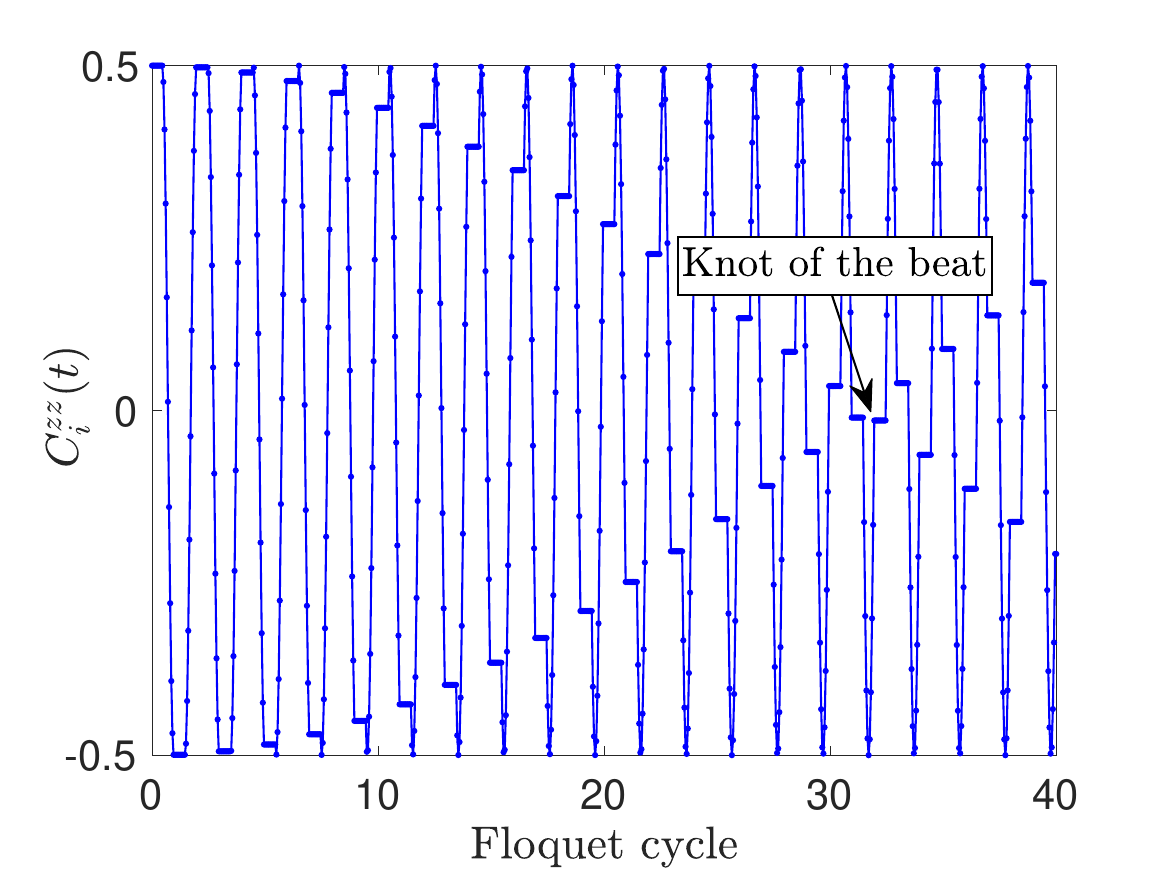}
 \caption{Effect of imperfect flip \new{($\epsilon=0.5$)} on a single uncoupled spin. One can see the emergence of a beating pattern along with the associated spikes leading to the formation of a knot at around Floquet cycle 32.}
 \label{uncoupled}
\end{figure}

%\begin{enumerate}
\emph{Case (i):} Uncoupled spins ($J=0, h=0$): This is the case when only the second part of the Hamiltonian 
(\ref{Hmain}) acts on the spins. Clearly, the spins stay uncoupled. While the system exhibits signatures such as period doubling as well as stability to infinite times, it cannot be called a time crystal because it is extremely sensitive to perturbations and is therefore, not a well defined phase of matter. This can be tested by introducing an imperfection in the spin flip denoted by $\epsilon$. This is shown in Fig.\ \ref{uncoupled} \new{ for $\epsilon=0.5$}.
Note that this regime is reminiscent of the setting of equal-position correlation function in independent two-level systems spreading over space as considered in Ref.\ \cite{PhysRevLett.114.251603}
as rather trivial and not time-crystalline behaviour.

We plot the equal space correlator defined in c
Eq.\ (\ref{eqspacecorrelator}) 
as a function of the Floquet cycle and from this we can see the beating pattern emerging and the knot of the beat at around the Floquet cycle 32. We also notice the spikes in the $C_i^{zz}(t)$ owing to the effect of non-zero $\epsilon$. We resort to looking at identifying signatures of time crystal from the short time dynamics as different values of $\epsilon$ would simply mean a change in the position of such knots along the X-axis. In other words, for $\epsilon=0$, the knot would completely disappear along with the spikes seen here. 

\emph{Case (ii):} Interacting spins in the absence of disorder ($J=1, h=0$): This is the case when both parts of the Hamiltonian in (\ref{Hmain}) are active, however, $h=0$ for $H_{\rm MBL}$. We will show the plots for $\epsilon =0,0.5$ starting from the N{\'e}el states as well as the polarized states.
     \begin{figure}
  \includegraphics[width=0.5\textwidth]{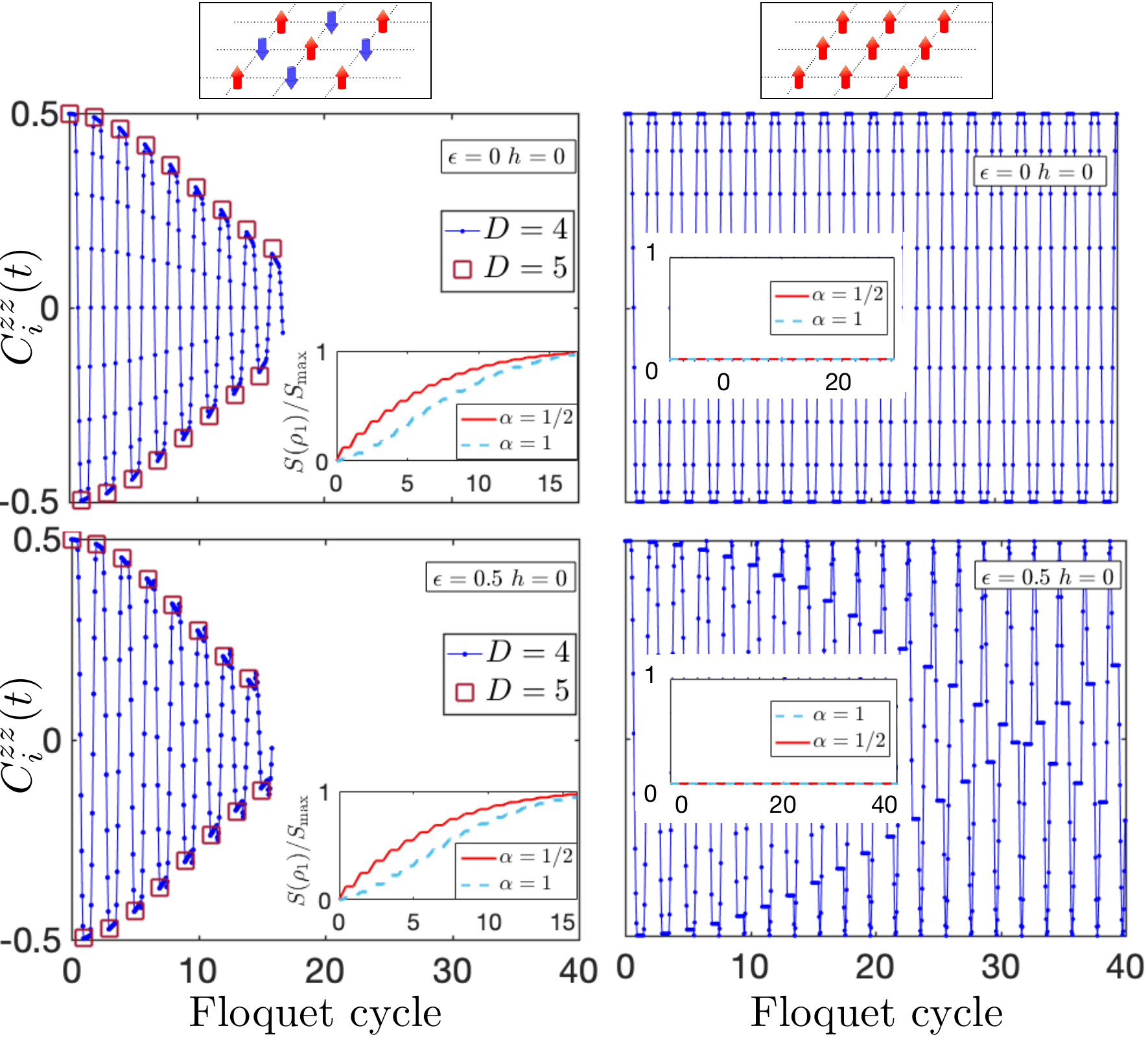}
  \caption{Equal space correlator evaluated at different times for coupled spins without disorder starting from N{\'e}el state (left) and polarized state (right) for the perfect (top) and imperfect flips (bottom). Time crystalline behaviour cannot survive for the N{\'e}el state due to thermalization while there is no dynamics for the initial polarized state. Also shown are simulations with $D=5$ for stroboscopic times. (Insets) growth of local R{\'e}nyi entropies $(\alpha=1/2,1)$ for the blue and red curve, respectively.}
  \label{nodisorder}
\end{figure}
In the presence of many-body interactions without any disorder, our system is 
expected to thermalize very quickly \cite{1408.5148,ngupta_Silva_Vengalattore_2011}. 
\new{An instance of a time evolution compatible with} this expectation is shown by plotting the equal space correlator in time in the left panels of Fig.\ \ref{nodisorder} where we are able to show time evolution 
only up to approximately 16 Floquet cycles. It is also evident from the growth of local R{\'e}nyi entropies for one-site reduced quantum states shown in the insets of the left panels. The R{\'e}nyi entropies $(\alpha=1/2,1)$ are re-scaled to their maximal value of $\log(d_p)$. \je{The values of $\alpha$ are chosen to provide an insight into how well one would expect a low bond dimension tensor to approximate the quantum state at hand \cite{Schuch_MPS}.}

The right panels of this figure correspond to starting from a polarized initial state and\new{, besides the periodic flipping of the spins induced by $H_S$,} there is no dynamics whatsoever which is also revealed by the local R{\'e}nyi entropies which stays zero throughout the evolution (shown in the insets). \new{ This can easily be understood by the spin SU(2) symmetry of the Hamiltonian $H_{\rm MBL}$ at $h=0$ which entails that even at non-zero $\epsilon$ the time evolution rotates all spins in the same manner during the application of $H_S$, while the spins remain inert during $H_{\rm MBL}$. After all, the fully spin-polarized state remains fully polarized (albeit with respect to a different polarization axis at $\epsilon\neq 0$) after each flip and the dynamics are the same as that of a single spin (compare Fig.~\ref{uncoupled})}.

 Even in such 
 \new{a short time}
 scale available (for the N{\'e}el initial state), one can 
 \new{still}
 see 
 \new{important}
 features such as period doubling
and the absence of spikes due to finite $\epsilon$, as seen in Fig.\ \ref{uncoupled}. One can then speculate that the destruction of the time crystal behaviour in this case is \new{ultimately} caused by thermalization
\new{for long times}. We then resort to overcoming this \new{effect} by adding strong disorder $h=100$ to our system for different levels of disorder $d_A=2$ and $d_A=5$.

\begin{figure}
  \includegraphics[width=0.5\textwidth]{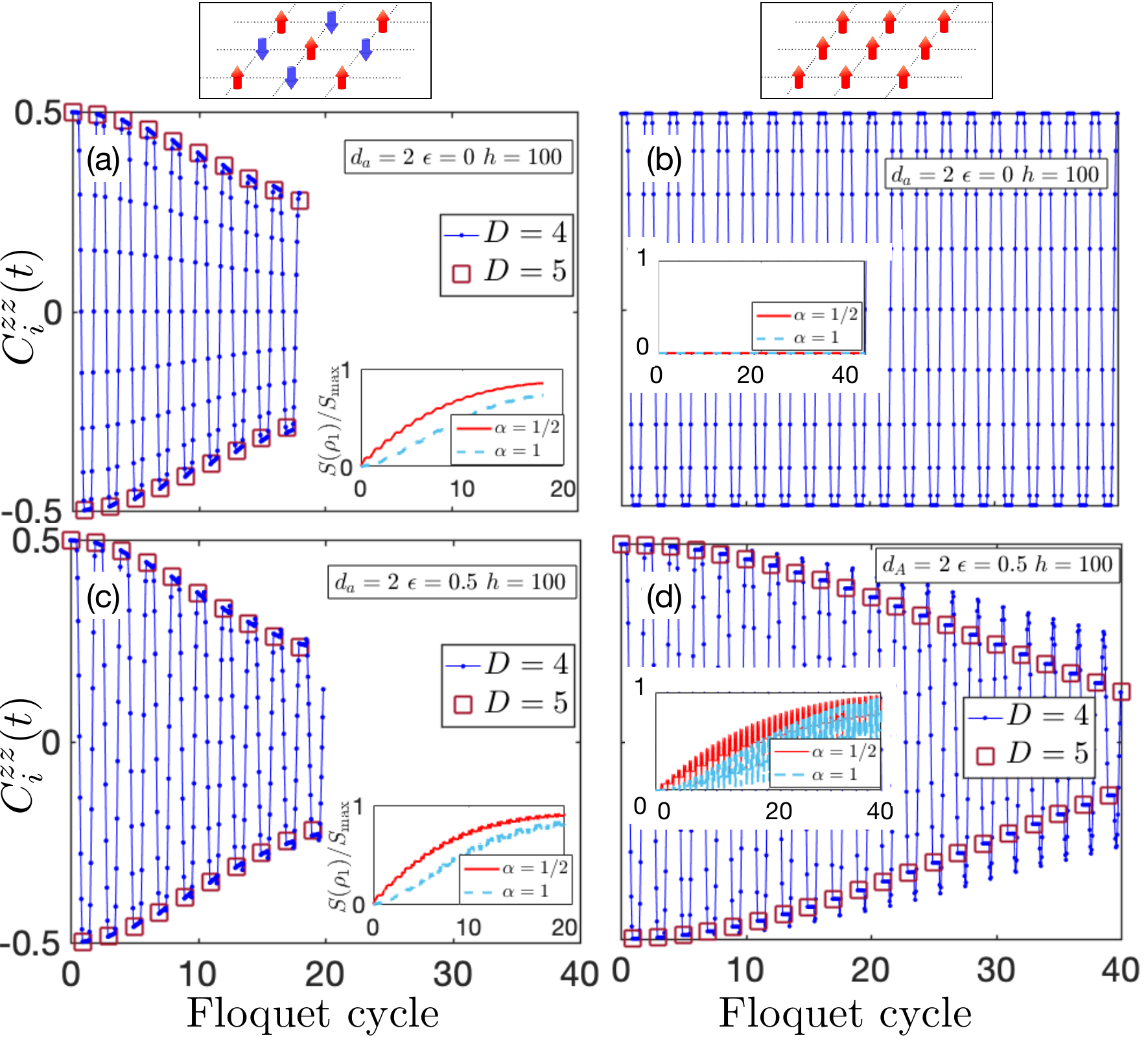}
  \caption{Equal space correlator evaluated at different times for coupled spins with strong disorder of two levels for different initial states and $\epsilon$. Again also shown are the simulations with $D=5$ for stroboscopic times. (Insets) growth of local R{\'e}nyi entropies $(\alpha=1/2,1)$ for the blue and red curve, respectively.}
  \label{strongdisordera2}
\end{figure}

\emph{Case (iii):} Interacting spins with strong disorder ($J=1$, $h=100$).
This is the case when both parts of the Hamiltonian are active and also the $H_{\rm MBL}$ \new{features}
strong disorder. We start our discussion by fixing the number of levels of disorder to 2. This means the disordered field can take only two different values. We see that, compared to case (ii), we have been able to delay thermalization only slightly for both $\epsilon=0$ and $\epsilon=0.5$ for the N{\'e}el initial states (Fig.\ \ref{strongdisordera2}). For these cases, we see that despite a strongly disordered field, the two levels are not enough to sufficiently stabilize a time crystal. This is consistent with our previous findings on many-body localization in two spatial dimensions \cite{Augustine2DMBL} where sufficient levels of disorder were required to achieve localization. However, if we start from the polarized state, we notice that the dynamics slow down considerably for the case of $\epsilon=0.5$ while there is none in the case of $\epsilon=0$. The local R{\'e}nyi entropies shown in the insets provide further substance to these results.  

\new{In a next step, we} now increase the number of levels of disorder by taking $d_a=5$ such that our disordered field can take five different values. In all the cases, we see that the stability of the time crystal improves remarkably, thereby allowing us to go much longer times even beyond the currently shown 40 Floquet cycle.

\begin{figure}
  \includegraphics[width=0.5\textwidth]{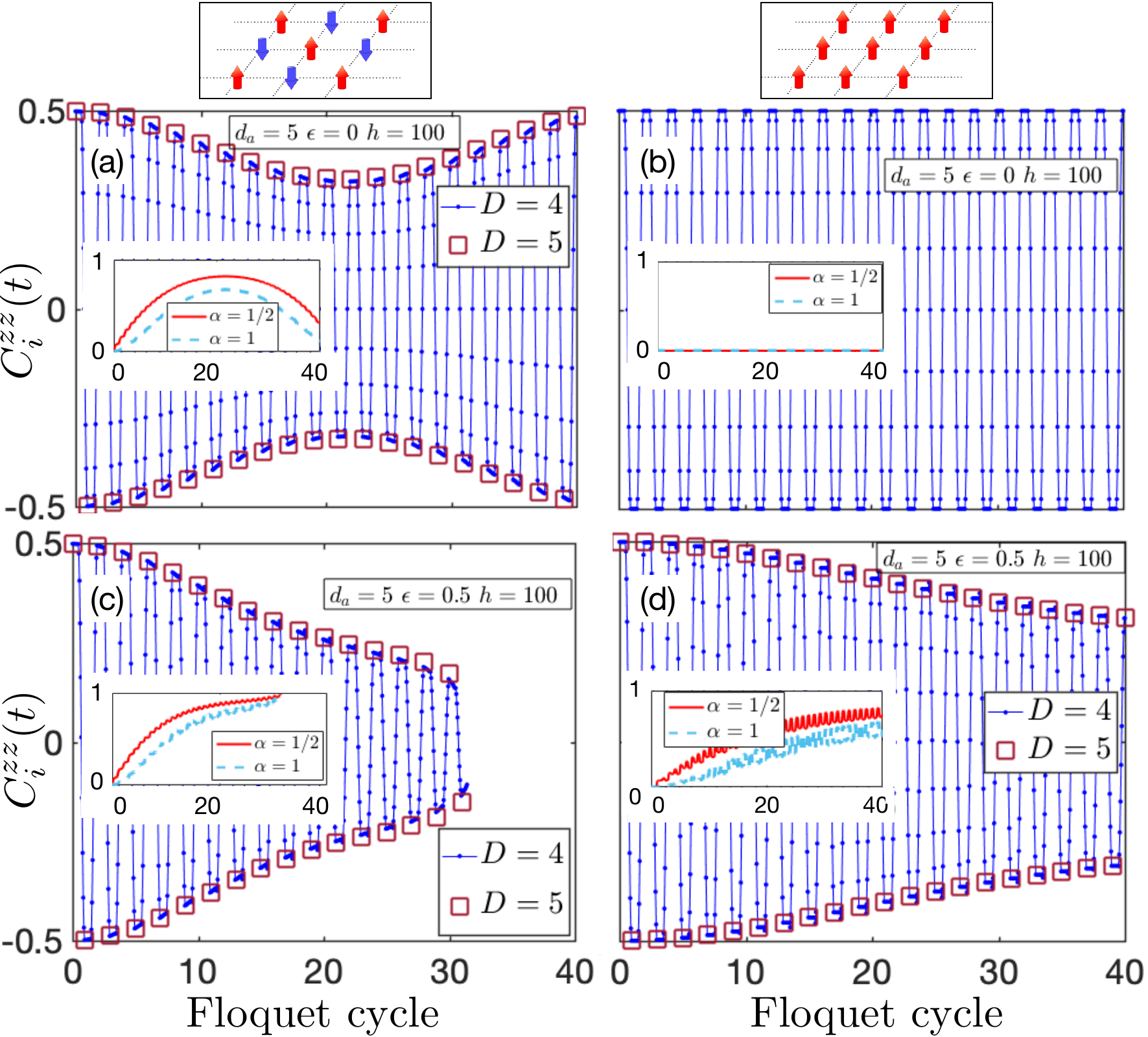}
  \caption{Same as Fig.~\ref{strongdisordera2}, but with five levels of disorder. The time crystal has stabilized due to the increasing number of disorder levels, thereby surviving much longer times for both the initial states.}
  \label{strongdisordera5}
\end{figure}
\begin{figure}
  \includegraphics[width=0.4\textwidth]{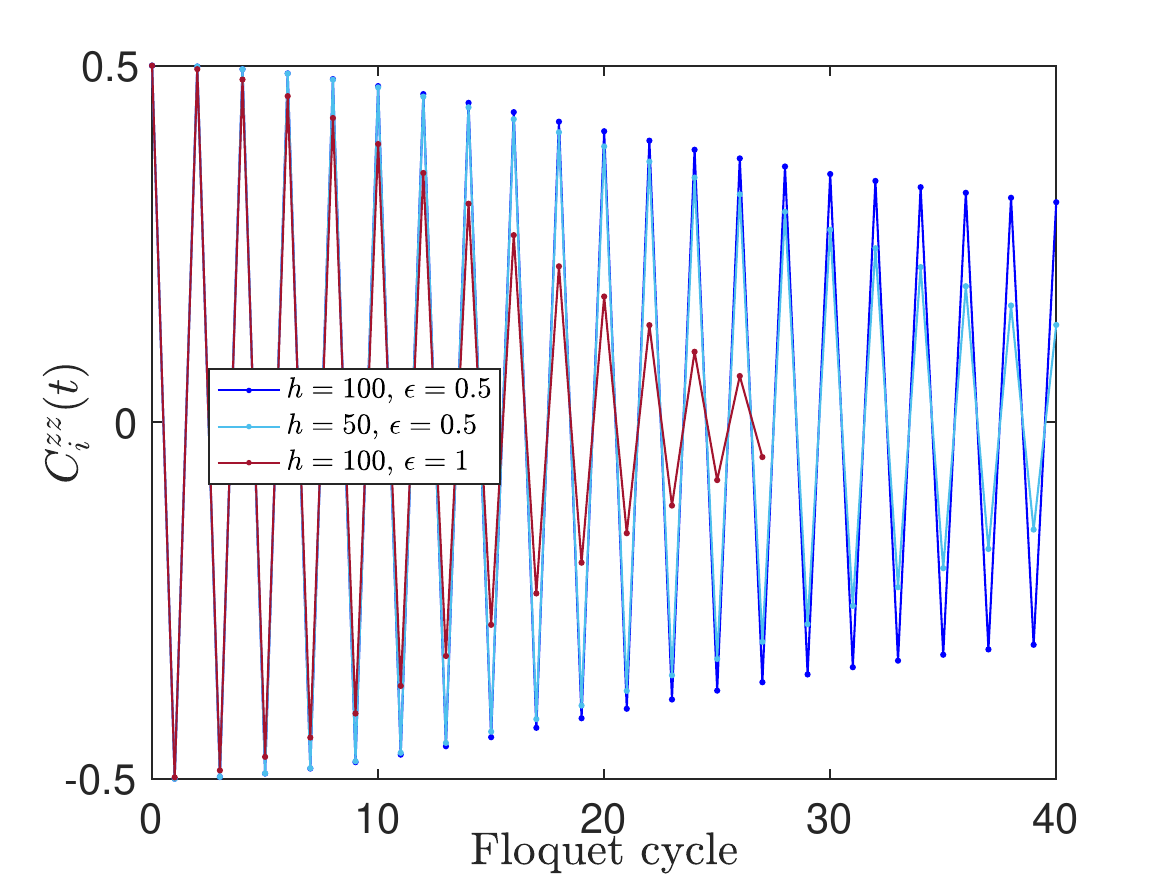}
  \caption{Stability of the time crystal with respect to different disorder strength $h$ and perturbation to the perfect flip $\epsilon$. The results are plotted only for stroboscopic times.}
  \label{robustnessTC}
\end{figure}

The delay of thermalization is also consistent with the growth of local R{\'e}nyi entropies (shown in the insets). Thus, we have managed to stabilize our two-dimensional time crystal (i.e., increase its lifespan) by increasing the number of levels of disorder at sufficiently large strength. We have also seen that the stability also depends on the initial state chosen, the one starting from polarized state being the most favourable. \new{ It is worth pointing out that a resurgence is observed in the top left panel of Fig.~\ref{strongdisordera5} due to the strong disorder and the perfect flip at $\epsilon=0$}. In Fig. \ref{robustnessTC}, we show the stability of our time crystal starting from the polarized state with respect to a smaller disorder ($h=50$) and a larger perturbation to the perfect flip ($\epsilon=1$). As expected, smaller disorder strength leads to a decrease in the lifespan of the time crystal. Unlike the case in Ref.\  \cite{Augustine2DMBL}, where only a modest disorder strength of $h=4$ (with at least four levels of disorder) was enough to localize our system, larger disorder strength is needed here in order to counter the extra heating effect of the drive. Similarly, a large value of $\epsilon$ will also lead to destruction of the time crystal ultimately~\cite{ZhangNature2017}. This is also shown in our Fig.\ \ref{robustnessTC}. \new{This behaviour is}
consistent with the findings of Ref.~\cite{ZhangNature2017}. 

\section{Implementations in programmable quantum simulators} 

It is key 
to the setting laid out here that it is amenable
to quantum simulation using programmable 
disorder. Dichotomic disorder seems incapable of
sufficiently stabilizing the time crystal, so that more intricate disorder is required. Specifically suitable for this kind of programmable disorder
seem systems of \emph{superconducting
systems} \cite{Dimitris, GoogleSupremacy} in which
two-spatial dimensions with programmable
disorder seem within reach, and also
discrete disorder can be largely programmed. Also, individually controlled 
\emph{Rydberg atoms} \cite{Browaeys}
and systems of \emph{cold atoms in optical 
lattices} \cite{2dMBLScience}
in which the disorder is realized 
by imaging a two-dimensional random disorder potential to a single atomic plane in an optical lattice, controlled by a \emph{digital mirror device} allowing for a programming of discrete disorder,
or possibly also by making use of another
atomic species acting as a disorder
potential to the other.
Similarly, two-dimensional
arrays of \emph{trapped ions} 
\cite{2DIons} 
have programming capabilities of the kind
required here. 

The situation discussed here gives rise to a highly interesting state of affairs, \new{linking notions of quantum many-body and quantum
information theory,}
to make a point mentioned in the
introduction more precise and specific: For short times,
\new{state-of-the-art} tensor network methods provide a reliable machinery to assess the pre-thermal time crystalline behaviour.
Based on such classically available data, \new{and pushing the classically accessible regime with substantial effort,} one can judge what phases of matter are \new{reasonably} to be expected, \new{with substantial predictive power}. 

Still, due to the 
linear growth of the entanglement entropy of subsystems in time \cite{SchuchQuench}, intermediate 
and long times \new{will remain inaccessible using tensor network methods, even if folding methods \cite{Folding} or the exploitation of mode transformations \cite{LongTimes} -- ideas that both have not yet been developed for two spatial dimensions -- may ultimately render data on somewhat longer times classically available.}

\new{This insight gives substantial impetus to the field of quantum simulation.
It} is perfectly compatible with complexity theoretic insights.
When appropriately cast into the form of a decision problem, the problem 
of simulating the time evolution under general translationally invariant local Hamiltonians  is
in the complexity class ${\tt BQP}$, the class of problems that can be solved by a quantum computer to 
bounded error in polynomial time. Interestingly, this problem is even complete for ${\tt BQP}$ \cite{Vollbrecht}, which
implies that it is unreasonable to expect that there exists an efficient classical algorithm for the problem, as this would imply that ${\tt BPP} = {\tt BQP}$, which would in turn result in polynomial time classical algorithms for factoring, which is considered excessively implausible. \new{Of course, physically ``natural'' problems may in principle outside the classically hard instances, but in the light of these complexity theoretic obstructions, it is still highly unlikely that a classical efficient simulation method for interacting quantum many-body systems can ever be identified.}

\new{This comes along with advice concerning the necessity of implementing quantum simulators: Indeed,}
one does
need programmable quantum simulators 
\cite{GoogleSupremacy,MonroeTimeCrystal,FlammiaProgrammable,Browaeys,Dimitris,ProbingQuantumSimulator,2dMBLScience,2DIons}
to ultimately assess the stability of such phases of matter for long times 
\cite{SondhiReview,doi:10.1146/annurev-conmatphys-031119-050658}. This 
constitutes a most interesting situation and at the same time a valid and meaningful application for programmable
quantum simulators as intermediate quantum technological devices between full quantum
computers and analog quantum simulators: Here, the long-time behaviour only accessible for quantum but not
classical devices is at the heart of the matter when discussing time-crystalline behaviour.

\section{Conclusion and outlook} 
In this work, 
we have provided fresh evidence of (prethermal) quantum time crystals in two
spatial dimensions, using the infinite version of projected entangled pair states (iPEPS)
that is able to address the thermodynamic
limit directly. We have combined a 
tensor network machinery that includes 
a quantum dilation approach to capture strong disorder with a suitable stroboscopic 
Floquet Hamiltonian evolution that 
features a discrete time translation symmetry. Starting from different initial states, we have clearly encountered the breaking of time translation symmetry revealed by the equal space correlator in time that is robust to perturbations in the spin flip Hamiltonian.  The stability of such 
quantum time crystals can be increased by taking a sufficiently strong disorder featuring a large number of levels of disorder and suitable initial state. In future work, we will explore whether dissipation can be properly engineered to stabilize such phases of matter by avoiding thermalization through the use of \emph{projected entangled pair operators} \cite{Kshetrimayum}.  

In addition to pushing the machinery of tensor networks to a new regime, it is the hope that 
this work provides significant
further guidance for the use of programmable quantum simulators to explore intricate non-equilibrium quantum phases of matter, after all to discriminate 
pre-thermal from genuine time crystals 
\cite{SondhiReview} as they cannot
classically discriminated. \new{Quantum simulators will also help to answer more intricate questions such as the one of broken translation symmetry in space as well as analyzing $C_{i,i'}^{O,O}(t,t')$ at $t'\neq 0$, which is more challenging to compute in our classical simulation approach. This could allow to obtain deeper insights into the mechanisms of spontaneous symmetry breaking in space and time \cite{PhysRevB.102.195116}.}

After all, phases such as discrete time crystals
may have an impact also beyond academic interests to devising technological applications in, say, 
quantum metrology. The present scheme adds to the portfolio of schemes for programmable quantum simulators to be explored that are feasible instances of quantum devices intermediate between
analog quantum simulators and full-scale quantum computers. \new{The present work, so we hope, also helps delineating
the delicate boundary between classically simulatable regimes of quantum many-body physics and those that ultimately have to be assessed with quantum devices.}

\section{Acknowledgements} 
 This work has been supported by the  Templeton Foundation, the FQXi, and the DFG 
 (EI 519/14-1, EI 519/15-1,  CRC 183 Project
 B01 and A03, and FOR 2724) and MATH+.
 This  work  has  also  received  funding  from  the  European  Unions  Horizon2020  research  and innovation  programme  under  grant  agreement  No.\  
 817482 (PASQuanS) on programmable quantum simulators. DMK acknowledges support under Germany's Excellence Strategy - Cluster of Excellence Matter and Light for Quantum Computing (ML4Q) EXC 2004/1 - 390534769 and from the Max Planck-New York City Center for Non-Equilibrium Quantum Phenomena.

% For Nat. Comm.
\section{Appendix} 

We make use of several tensor network methods 
here, developed to be applicable in new regimes, in particular a method rooted in \emph{infinite projected entangled pair states} (iPEPS)
\cite{iPEPSOld,SierraOld,NishinoOld}, \new{here pushed to the fresh regime of interactions, disorder and time-dependent drive being present simultaneously.} We implement disorder in such a translationally invariant setting using a quantum dilation technique~\cite{CiracdisorderTI}. We discuss briefly below, the different numerical methods used 
\new{and developed} 
in this work.

\subsection{Implementing disorder in a translationally invariant system} 
\begin{figure}
  \includegraphics[width=0.3\textwidth]{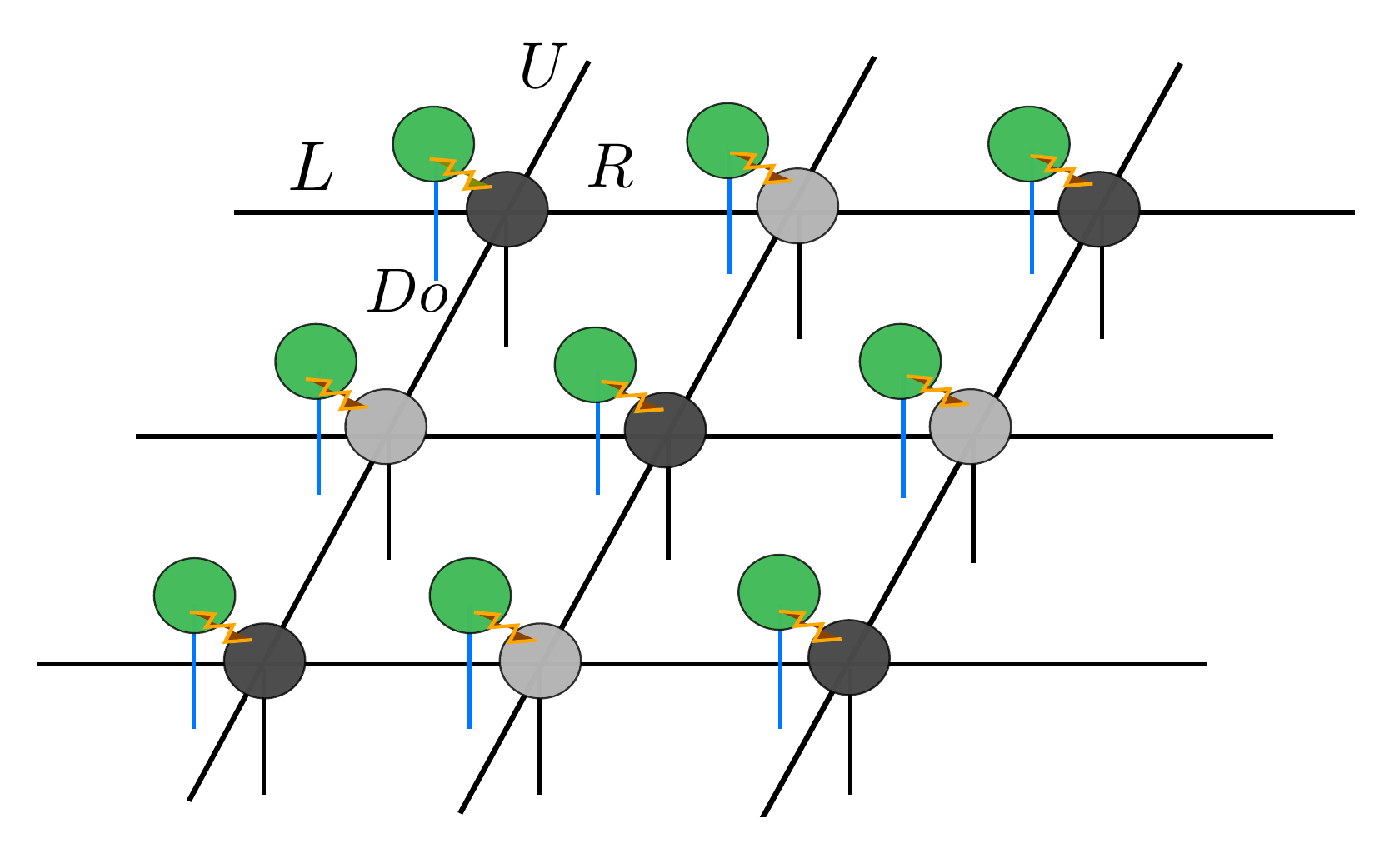}
  \caption{Implementing disorder in a translationally invariant setting. We choose our initial global state to be a tensor product of the physical state (represented by grey tensors)and the auxiliary state (represented by green tensors). $L$, $U$, $R$ and $Do$ correspond to the different links we need to update for a particular site.}
  \label{ipepsTC}
\end{figure}

This approach has originally been
introduced in Ref.\ \cite{CiracdisorderTI} for studying disordered infinite-sized one dimensional systems~\cite{Enss2017,SirkerPRL2014}. Such ideas have only very recently been implemented in two-dimensional systems using iPEPS~\cite{Claudiusevolution,Augustine2DMBL}. The procedure starts by appending an auxiliary site to each physical one in the system. The global initial state is thus a product state of the physical system and the auxiliary system represented by a state vector $|\Psi_0 \rangle = |\psi_{p_0} \rangle \otimes |\psi_{a_0} \rangle$. The auxiliary state is fixed and is a tensor product of equal superposition states at each site as defined in the main text.
The physical state depends on our choice and is also discussed in the main text. The original part of the Hamiltonian in 
Eq.\ (\ref{Hmbl}) is then modified to the form in Eq.~(\ref{Hanc}) which includes the classical interaction $S^z_{i_p}S^z_{j_p}$ between the physical sites and the auxiliary sites. This is responsible for injecting the disorder into the physical system. This is shown in Fig.~\ref{ipepsTC}. We then perform a quench of the global initial state with the Floquet Hamiltonian defined in Eq.\ (\ref{Hmain}), to arrive at
a time-dependent state vector
\begin{equation}
    |\Psi(t) \rangle = e^{-iH_Ft}|\Psi_0 \rangle .
\end{equation}
We note here that the different parts of the Hamiltonian act at different times of the evolution.

\subsection{Time evolution} 
In this work, we exploit a variant of \emph{time evolving block decimation} (TEBD)~\cite{VidalTEBD} in two spatial dimensions. We have specifically implemented an iPEPS algorithm with a two-site unit cell arranged in a checkerboard pattern. Such a configuration is enough to realize our two initial states discussed in the main text. The tensors are then optimized using the so-called \emph{simple update}~\cite{simpleupdatejiang}. This scheme is more efficient and has also been found to be more accurate and stable compared to the \emph{full update} procedure for real time evolution \cite{Claudiusevolution,Augustine2DMBL,Claudiuslocalquench}. To describe this approach briefly, consider our two-dimensional Hamiltonian defined on a square lattice as in the setting used here. We split our Hamiltonian into different parts 
\new{corresponding} to the different links of the lattice, i.e., 
\begin{equation}
H= H^L + H^U + H^R + H^{Do}
\end{equation}
where 
\begin{equation}
H^{\alpha} = \sum_{\langle i, j \rangle}h^{\alpha}_{i,j}
\end{equation}
for $\alpha=L,U,R,Do$ \new{correspond} to the `left',`up', `right' and `down' link of the square lattice. We then define the following two-body gate operators $e^{-i \delta t H} \approx e^{-i \delta t H^L} e^{-i \delta t H^U} e^{-i \delta t H^R} e^{-i \delta t H^{Do}}$. After updating each link with the appropriate term, the bond dimension along that link is truncated. The procedure is performed while assuming a mean-field like environment of the link. We use a first order Trotter step of $\delta t=0.005$. 

We have performed \new{such an} optimization procedure with an iPEPS bond dimension of $D=4,5$. The large physical dimension of the local Hilbert space ($d=d_p d_a =4,10$) restricts us from accessing very large bond dimensions $D$ of the iPEPS. This also limits the time scales that we one access using tensor network approaches (reflecting the observation that time evolution under local Hamiltonians is a computationally hard problem in worst-case complexity). The results shown above are, nonetheless, converged and consistent with the largest available bond dimensions used. Such an agreement between the highest available bond dimensions can be used to approximately determine the stopping criteria of our time evolution.

\subsection{Disorder-averaged expectation values}

Once the tensors have been obtained using the above procedure, we have resorted to computing the expectation values of the physical observables of interest. This involves contracting the entire two-dimensional tensor network in the thermodynamic limit. Unlike the  situation for one-dimensional matrix product states, this step cannot be done efficiently and is known to be a computationally hard problem
\cite{Contraction}, for exact
contraction even in average-case complexity \cite{AverageContraction}. Luckily, there are several approximation schemes available for this \cite{iPEPSOld,ctmnishino1996,ctmroman2009,Alexcontraction,hotrg,shi-jurancontraction} (and there is good evidence that the contraction of PEPS is computationally feasible to good approximation for physically meaningful PEPS describing gapped models 
\cite{PhysRevA.95.060102}. The expectation values of the observables computed here is already the disorder-averaged expectation value of all the possible disorder configurations. This is easy to see from the expression
\begin{eqnarray}
    \langle \langle \hat{O}(t) \rangle \rangle 
    &= &\langle \Psi_0|e^{iHt}\hat{O}e^{-iHt}|\Psi_0 \rangle\\
    &=& ( {}_a\langle \psi_0| \otimes {}_p\langle \psi_0|)e^{iHt}\hat{O}e^{-iHt}(|\psi_0 \rangle_p \otimes |\psi_0\rangle_a).
\nonumber
\end{eqnarray}
Thus, one of the advantages of our setting is having to avoid taking multiple shots of disorder and taking their averages. The programmable nature of the disorder employed here also has direct control over avoiding rare ergodic regions which can lead to instability of MBL at long times~\cite{deroeck2017,RoeckImbrie}

In this work, we
make use the \emph{corner transfer matrix renormalization group} method (CTMRG) \cite{ctmnishino1996,ctmnishino1997,ctmroman2009,ctmroman2012}, to approximate contraction, here freshly applied to classically
keeping track of quantum Floquet dynamics. This involves computing a set of tensors known as the \emph{environment} tensors approximating the fixed points of the ``corners'' of an infinite two-dimensional lattice. This infinite two-dimensional lattice to be contracted is actually composed of the double layer norm tensors of bond dimension $D^2$. We use an environment bond dimension of up to $\chi=D^2$ in our simulations which is found to be sufficient for our purposes. 

\begin{figure}
  \includegraphics[width=0.5\textwidth]{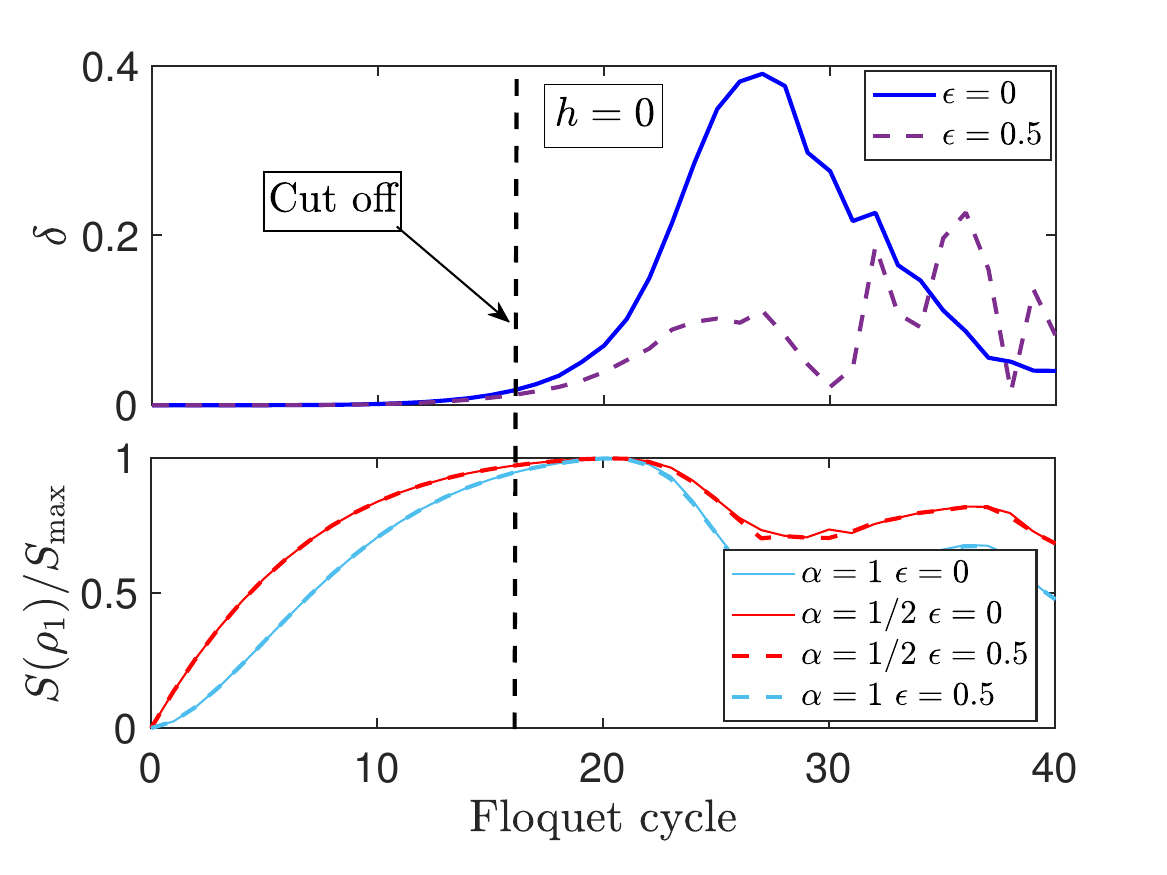}
  \caption{Errors corresponding to 
  Fig.\ \ref{nodisorder} (left panels). (Top) $\delta\geq 0$ is the absolute value of the difference of the expectation values of the correlators corresponding to simulations with $D=4$ and $D=5$. (Bottom) growth of local R{\'e}nyi entropies for one-site reduced density matrix for the full time evolution.}
  \label{error1}
\end{figure}

\begin{figure}
  \includegraphics[width=0.5\textwidth]{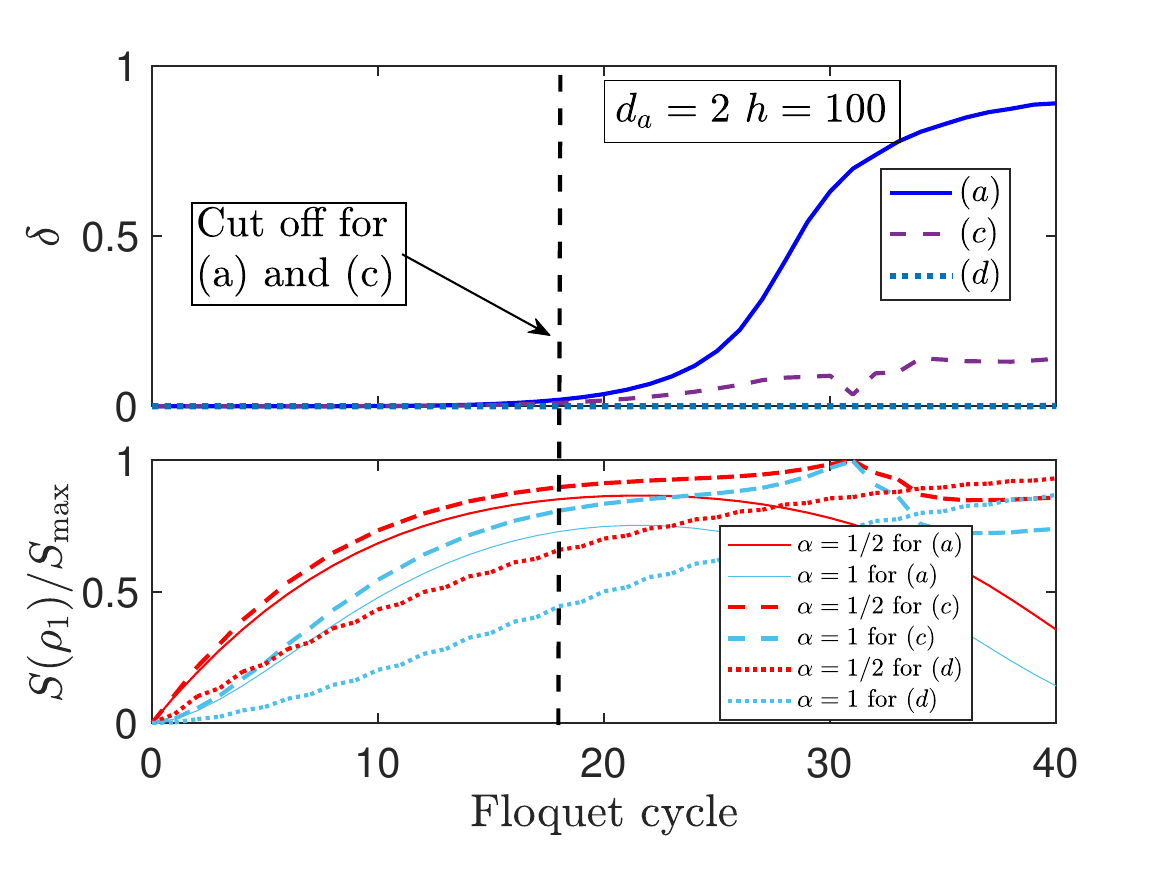}
  \caption{Errors corresponding to Fig.\ \ref{strongdisordera2}. (Top) $\delta$ is the absolute value of the difference of the expectation values of the correlators corresponding to simulations with $D=4$ and $D=5$. (Bottom) growth of local R{\'e}nyi entropies for one-site reduced density matrix for the full time evolution. $(a)$, $(c)$ and $(d)$ refer to the different subplots of Fig.\ \ref{strongdisordera2} in the results section.}
  \label{error2}
\end{figure}

\begin{figure}
  \includegraphics[width=0.5\textwidth]{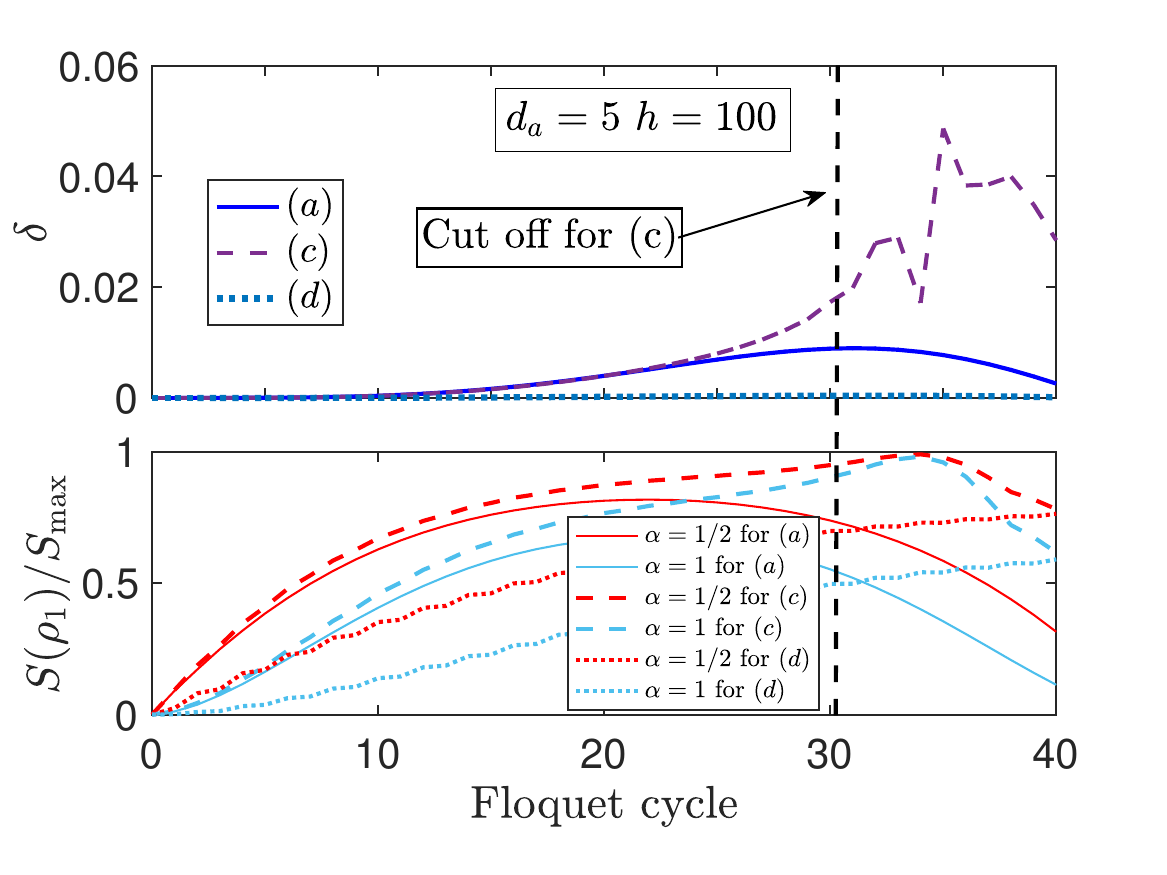}
  \caption{Errors corresponding to Fig.\ \ref{strongdisordera5}.
  (Top) $\delta$ is the absolute value of the difference of the expectation values of the correlators corresponding to simulations with $D=4$ and $D=5$. (Bottom) growth of local R{\'e}nyi entropies for one-site reduced density matrix for the full time evolution. $(a)$, $(b)$ and $(c)$ refer to the different subplots of Fig.\ \ref{strongdisordera5} in the results section.}
  \label{error3}
\end{figure}

%Incidentally, such a 'cut-off' agrees with the local R{\'e}nyi entropies saturating to it maximal value. 
In Fig.\ \ref{error1} (top panel), we plot $\delta\geq 0$, defined as the the absolute value of the difference between the expectation values of the correlators for bond dimensions $D=4$ and $D=5$.
In the bottom panel, we show the growth of local R{\'e}nyi entropies for the extended time evolution. The time at which $\delta$ becomes significant is used as the 'cut-off' time for Fig.\ \ref{nodisorder} (left panels). Incidentally, the local R{\'e}nyi entropies start saturating to its maximal value near this point. For the right panels of Fig.\  \ref{nodisorder}, there is no dynamics and therefore 
no concomitant errors in the growth of entanglement. Similar error measures corresponding to Fig.\ \ref{strongdisordera2} and Fig.\ \ref{strongdisordera5} are also shown in Fig.\ \ref{error2} and Fig.\ \ref{error3}, 
respectively. We see how the time evolution has been extended significantly depending on the choice of disorder levels and the initial states, directly corresponding to the stability of the time crystal.

\bibliographystyle{amsplain}
%\bibliography{BigReferences57}
\providecommand{\bysame}{\leavevmode\hbox to3em{\hrulefill}\thinspace}
\providecommand{\MR}{\relax\ifhmode\unskip\space\fi MR }
% \MRhref is called by the amsart/book/proc definition of \MR.
\providecommand{\MRhref}[2]{%
  \href{http://www.ams.org/mathscinet-getitem?mr=#1}{#2}
}
\providecommand{\href}[2]{#2}

\end{document}